\documentclass[twocolumn]{aastex631}

\usepackage{tabularx, booktabs}

\usepackage{mathtools}
\usepackage{savesym}
\savesymbol{tablenum}
\restoresymbol{SIX}{tablenum}
\usepackage{array}
\newcolumntype{P}[1]{>{\centering\arraybackslash}p{#1}}

\usepackage{booktabs}

\usepackage[utf8]{inputenc}
\usepackage{amsmath, amsthm, amssymb}
\usepackage{savesym}
\usepackage[italicdiff]{physics}
\usepackage{graphicx}
\usepackage[caption=false]{subfig}
\usepackage{cleveref}
\usepackage{layouts}

\shorttitle{Feedback and BH growth constraints with PTAs}
 \shortauthors{Tillman, Burkhart, Mingarelli et al.}

\graphicspath{{./}{figures/}}

\begin{document}

\title{Implications of the nanoHertz Gravitational-Wave Background for Galactic Feedback and Massive Black Hole Growth}
\author[0000-0002-1185-4111]{Megan Taylor Tillman}
\affiliation{Department of Physics and Astronomy, Rutgers University, 136 Frelinghuysen Rd, Piscataway, NJ 08854, USA}

\author[0000-0001-5817-5944]{Blakesley Burkhart}
\affiliation{Center for Computational Astrophysics, Flatiron Institute, 162 5th Avenue, New York, NY, 10010, USA}
\affiliation{Department of Physics and Astronomy, Rutgers University, 136 Frelinghuysen Rd, Piscataway, NJ 08854, USA}

\author[0000-0002-4307-1322]{Chiara M. F. Mingarelli}
  \affiliation{Department of Physics, Yale University, New Haven, CT 06520, USA}
  \affiliation{Center for Computational Astrophysics, Flatiron Institute, 162 5th Avenue, New York, NY, 10010, USA}
  
\author[0000-0002-5557-4007]{J. Andrew Casey-Clyde}
  \affil{Department of Physics, University of Connecticut, 196 Auditorium Road, U-3046, Storrs, CT 06269-3046, USA}
   \affil{Department of Physics, Yale University, New Haven, CT 06520, USA}

\author{Lars Hernquist}
    \affil{Center for Astrophysics | Harvard \& Smithsonian, 60 Garden St., Cambridge, MA 02138, USA}

\author{Sownak Bose}
    \affil{Institute for Computational Cosmology, Department of Physics, Durham University, South Road, Durham DH1 3LE, UK}

\author{Enik\H{o} Reg\H{o}s}
    \affil{HUN-REN CSFK Konkoly Observatory, Budapest, Konkoly Thege M 15-17}

\author{C\'esar Hern\'andez-Aguayo}
    \affil{LadHyX UMR CNRS 7646, Ecole Polytechnique, Institut Polytechnique de Paris, 91128 Palaiseau Cedex, France}

\begin{abstract}

We investigate how pulsar timing array (PTA) measurements of the nanoHertz gravitational-wave background (GWB) can constrain models for the growth history of supermassive black holes (SMBHs) and how active galactic nucleus (AGN) and stellar feedback models can affect GWB predictions. Feedback regulates supermassive black hole (SMBH) growth, altering the black hole mass function (BHMF).
Using BHMFs drawn from multiple cosmological simulation suites including IllustrisTNG, MillenniumTNG, Simba, and CAMELS, and combining these with a quasar-based SMBH binary population framework, we predict the resulting GWB amplitude under a range of different stellar and AGN feedback prescriptions.
 We find that the choice of both stellar and AGN feedback models alters the high-mass end of the BHMF and changes the predicted GWB amplitude by up to a factor of 2 for the fiducial simulations and a factor 10 for extreme feedback variations in CAMELS. Models with inefficient or absent AGN feedback produce abundant SMBHs and yield GWB amplitudes consistent with PTA data, yet fail in producing realistic galaxies. Fiducial models of AGN and stellar feedback suppress SMBH growth too much and under-predict the expected signal, an effect which could possibly be mitigated by more realistic black hole seeding and growth prescriptions. 
 The mismatch between the GWB amplitudes predicted by cosmological simulations and that inferred by PTA measurements suggests that SMBH growth is more efficient or occurs earlier than captured by current models. This demonstrates that PTA measurements provide a powerful new probe of not only the SMBH population but also feedback physics.

\end{abstract}

\keywords{Gravitational wave astronomy (675) --- Gravitational waves (678) --- Quasars (1319) --- Supermassive black holes (1663) --- Stellar feedback(1602) --- Active galactic nuclei(16) --- Cosmology(343)}

\section{Introduction}

Supermassive black holes (SMBHs) grow through accretion and mergers, and in doing so release enormous amounts of energy into their surroundings \citep{1963Natur.197.1040S,1978MNRAS.183..341W,2015ApJ...799..178K}. This process, broadly known as active galactic nucleus (AGN) feedback, regulates both the growth of SMBHs and the evolution of their host galaxies. Feedback can occur in radiative or kinetic modes, heating and expelling gas from galactic centers, suppressing further accretion, and altering star formation histories \citep{DiMatteo2005,Fabian2012,HeckmanBest2014,2023MNRAS.523.1104W}. Because AGN feedback controls the fueling of SMBHs, it leaves a direct imprint on the black hole mass function (BHMF) \citep{2025ApJ...990..120N}. Stronger feedback can truncate black hole growth, while weaker feedback allows more massive SMBHs to form, shifting the shape and normalization of the BHMF \citep{2015MNRAS.448.1504S, Sunseri:2025}.

The recent measurements by pulsar timing arrays \citep[PTAs,][]{NANOGrav:2023gor, EPTA:2023fyk, Reardon:2023gzh, Xu:2023wog, Miles:2024seg} provide strong evidence for the existence of a low frequency gravitational wave background (GWB), a signal that is likely generated by the incoherent superposition of SMBH binaries across cosmic time.
The origin of the signal is further implied to be produced by an astrophysical source, such as SMBH binaries, rather than a cosmological one given the agreement of point source astrophysical simulations with the theoretical expressions for the Fourier bin variances and the correlation of pulsar timing residuals \citep{Bernardo:2026}.
This would make the GWB a sensitive probe of the overall SMBH population in the Universe and the binary merger rate \citep{Kelley:2024}.
Its amplitude depends sensitively on the mass of the SMBHs, and therefore on the underlying BHMF~\citep{Phinney2001, Sesana2008}. \citet{CaseyClyde2022} made this connection explicit by building a quasar-based SMBH population model tied to an empirically constrained BHMF. They showed that the GWB amplitude can be decomposed into contributions from the BHMF, the number density of binaries, and the cosmological volume. In this framework, any modification to the BHMF directly propagates into a change in the predicted GWB amplitude. 

Since AGN feedback regulates SMBH growth via accretion, it is one of the dominant physical processes shaping the BHMF, in addition to mergers. 
The work presented here examines how both AGN and stellar feedback models modify the BHMF and, by extension, the predicted GWB amplitude. 
Many previous studies have extracted SMBH merger rates from large-volume simulations and translated them into PTA-band predictions \citep{Marinacci:2025, Mingarelli:2026review}. For example, using the \textsc{Illustris} simulation, \citet{Kelley2017b} showed that the SMBH merger population naturally produces a nanoHertz background at levels within reach of PTAs, and \citet{Sykes2022} employed \textsc{MassiveBlack-II} to forecast the GWB from its SMBH merger catalog.
\citet{Kozhikkal2024} explored how the black hole–bulge mass relation and its redshift evolution, calibrated with multiple cosmological simulations, impact predictions for the GWB. 
The \textsc{Astrid} simulation follows the orbital decay and pairing of SMBHs self-consistently during the run, providing so-called ``on-the-fly'' black hole dynamics that reduce the need for subgrid merger prescriptions~\citep{Ni2022}. 
\citet{Chen2025} find that \textsc{Astrid} yields a GWB amplitude amounting to roughly half of the measured PTA signal, underscoring the sensitivity of the background to assumptions about SMBH mergers.
\citet{Wang:2025} explores implications for LISA detections in Astrid and the potential overlap of the most massive LISA detections and PTA sources.
\citet{Izquierdo-Villalba:2022, Izquierdo-Villalba:2024} use the L-Galaxies semi-analytical model on the Millennium simulations to explore the impact of SMBH growth on both PTA measurements and to explore implications for LISA detection rates.
\citet{Barausse:2023} also explores the recent PTA measurements as a testbed for predicting future LISA detections.
\citet{Bonoli:2025} expand on the L-Galaxies semi-analytical model (dubbed L-GalaxiesBH)  and explore what is required of SMBH seeding and growth models to satisfy both PTA and JWST constraints.
These works complement ours by highlighting how host-galaxy scaling relations influence merger rates and the expected GWB amplitude, and how predictions from different simulations can be explored given the recent PTA measurements.

Several of these studies demonstrate that cosmological simulations can predict a GWB of the correct order of magnitude, but they do not explicitly explore how galactic feedback feeds into the GWB amplitude through the BHMF.
We conduct such an analysis herein.
Using the methodology of \citet{CaseyClyde2022}, we conduct a broad comparison of the BHMF and predicted GWB amplitude of three different cosmological hydrodynamic models: Simba, IllustrisTNG, and MillenniumTNG and also explore extreme feedback parameter variations of these models using the CAMELS simulations. 
The paper is laid out as follows. 
In Section \ref{sec:simsoverview} we describe how the broad range of AGN feedback models from IllustrisTNG, MillenniumTNG, Simba, and CAMELS and how these give rise to different BHMFs. 
In Section \ref{sec:GWB}, we review how \citet{CaseyClyde2022} links the BHMF  to the GWB amplitude, and in Section \ref{sec:results} we report our results. We discuss our results in \ref{sec:discussion} and summarize our conclusions in Section \ref{sec:conclusion}.

\begin{figure*}[t!]
    \centering
    \includegraphics[width=\linewidth, trim={0 0 9cm 0},clip]{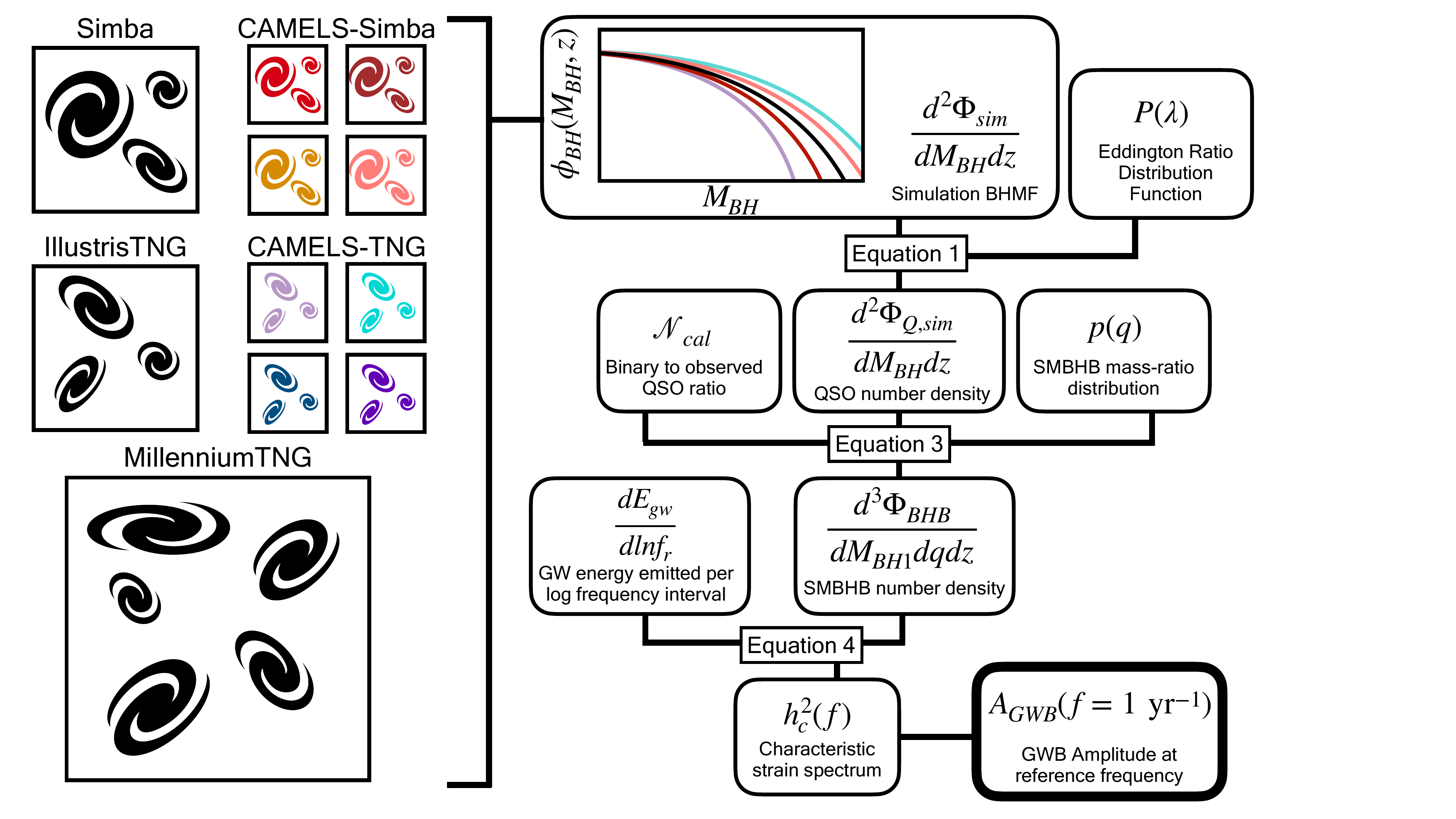}
    \caption{Graphic illustrating the pipeline used to obtain the GWB amplitude from the simulation BHMFs. We explore different simulation models (left-side), MillenniumTNG/IllustrisTNG and Simba, and variations of said models using the CAMELS simulation suites. These different simulations predict different BHMFs which are used in our semi-analytic model to calculate the characteristic strain spectrum (right-side), described in Section \ref{sec:GWB}.}
    \label{fig:graphic}
\end{figure*}

\begin{figure}[t!]
    \centering
    \includegraphics[width=0.8\linewidth]{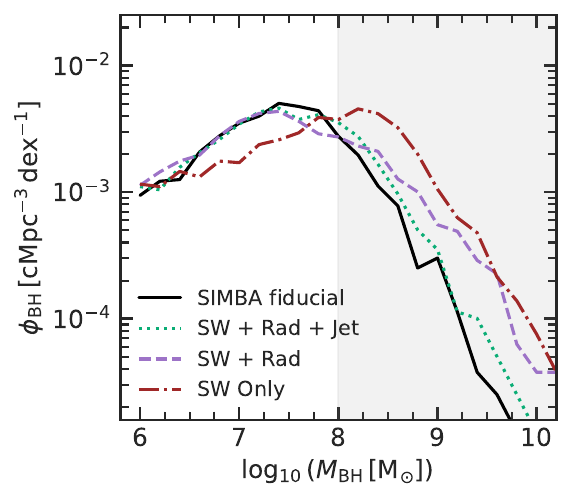}
    \caption{The $z=0$ BHMF predicted by the Simba simulation model when implementing different AGN feedback modes. The volume, resolution, cosmology, and all other astrophysics are the same in these simulations.  This illustrates how much variation can occur in the predicted BHMF due to AGN feedback. Variations on the high mass end ($M_\textnormal{BH} \gtrsim 10^8 M_\odot$, marked by the shaded grey region) of the mass function will most impact the GWB amplitude with higher mass SMBHs contributing more.}
    \label{fig:feedback}

\end{figure}

\section{Simulations Overview}
\label{sec:simsoverview}

    We explore the IllustrisTNG, MillenniumTNG, Simba, and CAMELS simulation suites and analyze how variations in the galactic feedback models impact the predicted BHMF.
    Differences in the predicted BHMF have implications for measurements of the stochastic GWB amplitude.
    For each suite, we also analyze a set of simulations with varying box sizes to conduct a convergence study. This is important because the predicted GWB amplitude is sensitive to the high mass end of the BHMF, which itself is sensitive to box size.
    In the following sub-sections, we describe the different cosmological hydrodynamic simulation models analyzed  and the variations of said models explored. 
    We also describe how we extract the BHMF from these simulations.

        \subsection{The IllustrisTNG and MillenniumTNG Simulations}
    
        The IllustrisTNG simulations are run using the AREPO code \citep{Springel:2010, Weinberger:2020} and model gravitational interactions through the TreePM algorithm \citep{Springel:2005}.
        Radiative cooling, including line cooling, free-free emission, and inverse Compton cooling, is modeled as described in \citet{Katz:1996}.
        Metal and metal-line cooling are described in \citet{Vogelsberger:2012, Vogelsberger:2013}.
        IllustrisTNG assumes ionizing equilibrium and a uniform ionizing background model, which accounts for on-the-fly hydrogen self-shielding \citep{Faucher-Giguere:2009, Rahmati:2013}.
        Star formation and the ISM sub-grid model are from \citet{Springel:2003}, and enrichment is modeled from Type Ia and II Supernovae (SNe), asymptotic giant branch (AGB) stars, and neutron star mergers.
        The IllustrisTNG simulations use the cosmological parameters: $\Omega_m = 0.3089$, $\Omega_\Lambda$ = 0.6911, $\Omega_b = 0.0486$, $H_0 = 67.74$ km/s/Mpc, $\sigma_8 = 0.8159$, and $n_s = 0.9667$.

        IllustrisTNG models galactic feedback from both stars and AGN.
        Stellar feedback is modeled as kinetic (with a thermal sub-component) galactic winds that are temporarily hydrodynamically decoupled.
        These winds are stochastically and isotropically ejected from star-forming gas. 
        This model is further described in \citet{Pillepich:2018a}.
        
        In IllustrisTNG, SMBHs are seeded with mass $M_\textnormal{BHseed} = 8\times10^{5}M_\odot/h$ in halos with mass $M_\textnormal{FoF} > 5 \times 10^{10} M_\odot /h$. BH accretion in IllustrisTNG uses the \citet{Bondi:1952} prescription, capped at the Eddington limit. 
        AGN feedback includes three modes: thermal, kinetic, and radiative.
        The radiative mode is always on and adds the radiation flux of the SMBH to the cosmic ionizing background, heating the gas nearest to the host halo. 
        Which of the other two modes is active depends on the accretion rate of the BH.
        At high accretion rates, the thermal feedback mode is active, while at lower accretion rates, the kinetic feedback mode is active. 
        For both of these modes, energy is injected into the `feedback sphere' of the SMBH.
        Unlike stellar feedback, AGN feedback in IllustrisTNG is never decoupled hydrodynamically.
        For more information on the IllustrisTNG model, we refer the reader to \citet{Pillepich:2018a} and \citet{Nelson2019}.

        We analyze three different IllustrisTNG simulations, each with different box sizes but all implementing identical physics.
        The smallest box size has a co-moving side length of 35 Mpc$/h$ with $2160^3$ gas particles and an initial mass resolution of $8.5\times10^4 M_\odot$ (TNG50).
        The next box has a side length of 75 Mpc$/h$ with $1820^3$ gas particles and an initial mass resolution $1.4\times10^6 M_\odot$ (TNG100).
        The largest box has a side length of 205 Mpc$/h$ with $2500^3$ gas particles and an initial mass resolution $1.1\times10^7 M_\odot$ (TNG300).
        We use these different box sizes to explore convergence of the BHMF.

        We also utilize the MillenniumTNG simulation to test the convergence of the BHMF. 
        The MillenniumTNG simulation uses identical physics and parameters to IllustrisTNG, with only a few minor modifications where small issues found during and after the IllustrisTNG run were fixed \citep{Pakmor:2023}.
        The MillenniumTNG simulation has a co-moving box side length of 500 Mpc$/h$ with $4320^3$ gas particles and an initial mass resolution of $3.1\times10^7M_\odot$ (MTNG740).
        We discuss the impact of box size on predictions for the GWB amplitude in Section \ref{sec:discussion}.

    \subsection{The Simba Simulations}

        The Simba simulations are run with GIZMO's meshless finite mass hydrodynamics \citep{hopkins:2015} and employ state-of-the-art subgrid physical processes to form realistic galaxies \citep{dave:2019}.
        Based on GADGET-3 \citep{Springel:2005}, the GIZMO gravity solver evolves dark matter and gas simultaneously with gravitational and pressure forces.
        It models shocks via a Riemann solver with no artificial viscosity.
        Radiative cooling and photoionizing heating are modeled with Grackle3.0 which includes non-equilibrium evolution of primordial elements, a partially-uniform ionizing background, hydrogen self-shielding, and metal cooling \citep{Smith:2017, Haardt:2012, Rahmati:2013}. 
        The Simba simulations utilize cosmological parameters: $\Omega_m = 0.3$, $\Omega_\Lambda$ = 0.7, $\Omega_b = 0.048$, $H_0 = 68$ km/s/Mpc, $\sigma_8 = 0.82$, and $n_s = 0.97$.

        The Simba simulations model both stellar and AGN feedback.
        Simba tracks chemical enrichment from Type Ia SNe, Type II SNe and from AGB stars \citep{Li:2019}.
        Stellar feedback is modeled through kinetic galactic winds that are temporarily hydrodynamically-decoupled, two-phase, and metal-enriched. 
        The SNe feedback prescriptions are based on the FIRE zoom-in simulations \citep{hopkins+2014, Muratov:2015, DAA:2017}.
        
        In Simba, SMBHs are seeded with mass $M_\textnormal{BHseed} = 10^4 M_\odot/h$ in galaxies that do not already contain a SMBH based on a stellar mass threshold $M_* \gtrsim 10^{9.5} M_\odot$. 
        BH accretion is based on both the gravitational torque and Bondi accretion models, with cold and hot gas accreting in different modes \citep{Bondi:1952, Hopkins+Quataert:2011, DAA:2017a}.
        Similarly, AGN feedback is modeled in two modes: with high accretion rate mass loading outflows (the radiative mode) and low accretion rate jets (the jet mode). 
        For all AGN feedback, gas is ejected in a bipolar fashion, parallel to the angular momentum vector of the gas in the BH kernel. 
        Accompanying the AGN jets, that reach maximum allowed speeds, is an X-ray feedback mode that injects energy into the immediate surroundings of the BH.
        This additional feedback mode is added to capture the local effects of AGN jets since, due to resolution limitations, the AGN jets are temporarily hydrodynamically-decoupled.
        For more details on Simba, we refer the reader to \citet{dave:2019}.
    
        The flagship simulation has a comoving box length of 100 Mpc$/h$ and an initial gas element mass resolution of $1.82\times10^7$ $M_\odot$. 
        We also analyze a set of 50 Mpc$/h$ boxes with mass resolution $2.28\times 10^6 M_\odot$ that toggles the galactic feedback modes on and off.
        Of these smaller boxes, we analyze four simulations: a stellar feedback only simulation (`SW only'), one that adds in radiative AGN feedback (`SW + Rad'), one that adds in AGN jets (`SW + Rad + Jet'), and a simulation with the fiducial feedback implementation (`Fiducial').
        
    \subsection{The CAMELS Simulations}

        The Cosmology and Astrophysics with MachinE Learning Simulations (CAMELS) project consists of thousands of N-body and state-of-the-art hydrodynamic simulations that span a vast cosmological and astrophysical parameter space \citep{camels, CAMELS-public}.
        Each simulation is run with $256^3$ gas particles in a box with co-moving side length $25$ Mpc$/h$ and an initial mass resolution of $1.27\times10^7M_\odot h$.
        For this work, we analyze the simulations run with the IllustrisTNG and Simba models, which all use the same cosmological parameters: $\Omega_m = 0.3$, $\Omega_b = 0.049$, $H_0 = 67.11$ km/s/Mpc, $\sigma_8 = 0.8$, $M_\nu$ = 0 eV, $w=-1$, and $n_s = 0.9624$.
        
        The CAMELS-Simba and CAMELS-TNG simulation suites feature 28 parameter variations that explore both galaxy formation and cosmology \citep{Ni:2023}.
        Of these 28 parameters, 23 control the stellar modeling, SMBH modeling, and the implemented galactic feedback.
        These variations allow for an in depth exploration of which subgrid models can impact the predicted BHMF and, in turn, the GWB. 
        
        The CAMELS project includes a set of simulations, referred to as the 1P set, that vary parameters one at a time while keeping all other aspects identical, including the initial conditions.
        We analyze the 1P set to determine which subgrid physics plays a role in setting the predicted GWB.
        The 23 astrophysics parameters that are explored herein are summarized for the IllustrisTNG suite in Table \ref{tab:TNGsummary} and for the Simba suite in Table \ref{tab:SIMBAsummary}.
       Also included in the CAMELS project is a set of 27 simulations that are run with identical physics and cosmology but with varying initial conditions, referred to as the CV set.
        We analyze the CV set, alongside our box size analysis, to examine the impact of cosmic variance on the predicted GWB.  

        Larger box simulations are preferred for studying the GWB since convergence at the high mass end of the BHMF requires sufficient sampling of high mass halos, however larger box simulations are also more expensive to run.
        The benefit of having a large number of smaller box simulations through the CAMELS project is that we can isolate and explore the impact of individual astrophysics  parameters by comparing simulations that vary only that parameter.
        Determining what physics has an impact can motivate future works to explore those effects in larger box simulations where the high mass end of the BHMF is more robust.
        Figure \ref{fig:graphic} illustrates this idea and also summarizes the pipeline we use to calculate the predicted GWB amplitude for a simulation using its BHMF.

    \subsection{Calculating the BHMF}

    We calculate the BHMF by binning the SMBHs in each simulation into 0.2 dex bins. The BHMFs are calculated for discrete redshift snapshots from $z=0$ to $z=2$. These differential number densities, $\Phi(M_\mathrm{BH}, z)$, are then used directly in Eq. \eqref{eq:hc_integration} to account for the redshift evolution of the population.
    For the CAMELS-TNG and CAMELS-Simba simulations, we pull the SMBH population from 23 different snapshots in the redshift range of interest.
    For the Simba simulations we similarly pull the SMBH data from 23 different snapshots that were chosen to match the CAMELS redshifts.
    For IllustrisTNG and MillenniumTNG the SMBH populations are instead pulled from the subhalo catalogs.
    How the BHMFs predicted by the simulations are used in the calculation of the GWB amplitude is described in the next section.

\section{Quasar-based GWB Model}
\label{sec:GWB}

We adopt the SMBHB population framework of \citet{CaseyClyde2022} to transform simulation-predicted BHMFs into a GWB signal. In this approach, the empirically measured QLF is used as a tracer of the SMBHB population. This QLF-based model leverages observed quasar demographics to constrain the mass distribution of SMBHBs and predict the GWB amplitude.

\citet{CaseyClyde2022} computed the expected GWB, assuming that quasars are initially triggered by galaxy mergers. They found that for the model to reproduce the NANOGrav signal, the binary population must be approximately four times larger than the quasar population, implying a duty cycle where $\sim$25\% of binaries are hosted by quasars. By calibrating our simulated BHMF against this observed quasar population, we can utilize the framework in \citet{CaseyClyde2022, CaseyClyde2025} to obtain a self-consistent estimate of the GWB strain.

The transformation from the BHMF, from the cosmological simulations, to the BH Binary Mass Function (BHBMF) required for the GWB calculation proceeds in three steps: (1) isolating the active quasar population, (2) scaling to the total binary population, and (3) assigning secondary masses.

First, we treat the total BHMF from the simulation, $\phi_{\rm sim}(M_{\rm BH}, z)$, as the parent population. We derive the subset of these black holes that would appear as active quasars, $\phi_{Q, \rm sim}$, by convolving the parent mass function with an Eddington Ratio Distribution Function (ERDF), $P(\lambda)$. The ERDF effectively acts as a duty cycle, quantifying the probability that a black hole of mass $M_{\rm BH}$ is accreting at Eddington ratio $\lambda = L_{\rm bol}/L_{\rm Edd}$. The number density of simulated quasars is therefore:
\begin{equation}
\frac{d^2 \Phi_{Q, \rm sim}}{dM_{\rm BH}\,dz} = \frac{d^2 \Phi_{\rm sim}}{dM_{\rm BH}\,dz} \times \int_{\lambda_{\rm min}}^{\lambda_{\rm max}} P(\lambda)\,d\lambda,
\end{equation}
where the integral represents the duty cycle $f_{\rm duty}(M_{\rm BH}, z)$, or the fraction of black holes that are radiatively active \citep{CaseyClyde2025}. 

We then convert this active population into a binary population. Following the calibration in \citet{CaseyClyde2022}, we scale the active quasar population by a normalization factor $\mathcal{N}_{\rm cal} \approx 4$, reflecting the observational constraint that there are roughly four times as many binaries as there are visible quasars. This is a model-inferred calibration factor derived from fitting quasar-based models to GWB limits~\citep{CaseyClyde2022}, rather than a direct observational constraint.

The primary mass distribution of the binaries is therefore:

\begin{equation}
\frac{d^2 \Phi_{\rm BHB}}{dM_{\rm BH,1}\,dz} \approx \mathcal{N}_{\rm cal} \times \frac{d^2 \Phi_{Q, \rm sim}}{dM_{\rm BH,1}\,dz}.
\end{equation}
Finally, because we do not explicitly track binary pairings, we assign a secondary mass $M_{\rm BH,2}$ to each primary $M_{\rm BH,1}$ by sampling a mass ratio $q = M_{\rm BH,2}/M_{\rm BH,1}$. The full differential SMBHB number density is given by:
\begin{equation}
\frac{d^3 \Phi_{\rm BHB}}{dM_{BH,1}\,dq\,dz} \;\propto\; 
\left( \mathcal{N}_{\rm cal} \frac{d^2 \Phi_{Q, \rm sim}}{dM_{BH,1}\,dz} \right) \;\,p(q),
\label{eq:bhb_pop}
\end{equation}
where $p(q)$ is the SMBHB mass-ratio distribution, which we take as a log-normal centered at $q=0.33$ with width 0.5 dex. 
We adopt this distribution to remain consistent with the calibration of $\mathcal{N}_{\rm cal}$ in \citet{CaseyClyde2022}, though we acknowledge that theoretical models often predict log-uniform distributions favoring lower mass ratios.

Here $p(q)$ determines only the secondary mass; the primary mass $M_{\rm BH,1}$ is drawn directly from the simulation's feedback-regulated BHMF.

The characteristic strain spectrum $h_c(f)$ of the GWB is obtained by integrating over this derived binary population:
\begin{equation}
h_c^2(f) =\!\! \int \!\!\int\!\!\int \!\!\frac{1}{1+z}
\frac{d^3 \Phi_{\rm BHB}}{dM_{BH,1}\,dq\,dz}\;
\frac{dE_{\rm gw}}{d\ln f_r}\;
dM_{BH,1}\,dq\,dz,
\label{eq:hc_integration}
\end{equation}
where the term $dE_{\rm gw}/d\ln f_r$ is the GW energy emitted per logarithmic frequency interval. For circular binaries, the GW energy spectrum is \citep{Phinney2001}:
\begin{equation}
\frac{dE_{\rm gw}}{d\ln f_r} \;=\; \frac{\pi^{2/3}}{3}\,\mathcal{M}^{5/3}\,f_r^{2/3},
\label{eq:dEdf}
\end{equation}
where $\mathcal{M} = M_{\rm BH,1} q^{3/5} (1+q)^{-1/5}$ is the chirp mass and $f_r=f(1+z)$ is the rest-frame frequency. In this work, we insert the BHMFs predicted by cosmological simulations ($\Phi_{\rm sim}$) into this framework, thereby mapping theoretical SMBH growth models directly to the GWB amplitude.

Finally, we calculate the characteristic strain assuming a power-law spectrum $h_c(f) \propto f^{-2/3}$ (Phinney 2001). This approximation implicitly neglects environmental hardening mechanisms (e.g., stellar scattering) that may flatten the spectrum at low frequencies, as well as the finite number of sources that introduces anisotropies at high frequencies. We use this power law to report the characteristic GWB amplitude ($A_{GWB}$) at a standard reference frequency of $f=1\,\mathrm{yr}^{-1}$.

\begin{figure*}[t!]
    \centering
    \includegraphics[width=0.7\linewidth]{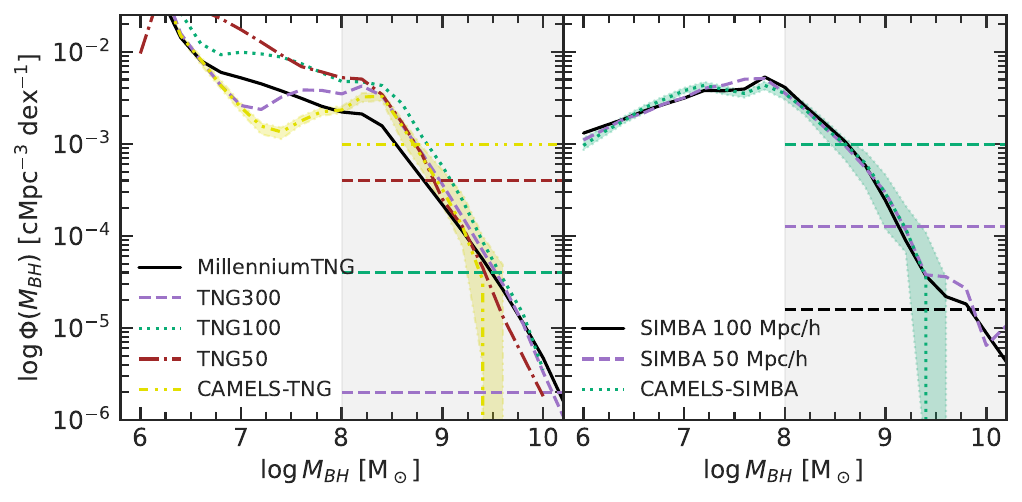}
    \caption{The BHMF (integrated from $z=0$ to 2) predicted by the IllustrisTNG (left) and Simba (right) simulations for different box sizes. MillenniumTNG, TNG300, TNG100, TNG50, and CAMELS-TNG/Simba are run in a box with co-moving side length 500, 205, 75, 35, and 25 Mpc/h respectively. The Simba simulations are run in a box with co-moving side length as labeled in the legend. 
    The dashed horizontal line corresponds to where mass bins contain fewer than 10 SMBHs matched to the simulations by color.
    Both simulation models struggle to produce a well sampled high mass ($M_\textnormal{BH} > 10^{9} M_\odot$) BHMF in a box with side length 25 Mpc/h, i.e.\ in the CAMELS simulations. These results imply that a box size length of at least $35$ Mpc/h for IllustrisTNG and $50$ Mpc/h for Simba is required to produce a reasonable sampled BHMF for $M_\textnormal{BH} > 10^{9} M_\odot$. For the CAMELS simulations, the shaded region corresponds to the 16th-to-84th percentiles from the CV set.}
    \label{fig:convergence}
\end{figure*}

\section{Results}
\label{sec:results}

Here we discuss the resulting BH mass functions and predicted GWB amplitudes from the IllustrisTNG, MillenniumTNG, Simba, and CAMELS simulations.

\subsection{AGN Feedback and the Black Hole Mass Function}

We first demonstrate the impact of altering AGN feedback on the BHMF. Using the Simba AGN feedback variation 50 Mpc/$h$ side length boxes, we compute the BHMF in four simulations: a stellar feedback only simulation (`SW only'), one that adds in radiative AGN feedback (`SW + Rad'), one that adds in AGN jets (`SW + Rad + Jet'), and a simulation with the fiducial feedback implementation (`Fiducial').
Figure \ref{fig:feedback} shows the predicted BHMF at $z=0$ from these simulations to illustrate how much the mass function can vary given different galactic feedback implementations, motivating further exploration of this effect. The Simba fiducial model (black line) predicts significantly fewer massive black holes that would fall in the PTA band than the other models, which remove different aspects of the AGN feedback model. Removing some or all aspects of the AGN feedback (colored dashed/dotted lines in Figure \ref{fig:feedback}) results in more high-mass black holes, and, as shown in many previous works, produces simulations with unrealistic galaxy properties and intergalactic medium properties \citep{Christiansen:2020, Tillman:2023, Tillman:2025}.

\subsection{The SMBH mass function and convergence}

For each BHMF plot, we highlight the $M_\textnormal{BH}\geq10^8M_\odot$ mass range since it is the most relevant for predictions of the GWB amplitude.
However, even within this range the higher mass BHs will contribute more to the predicted amplitude.
We calculate the BHMF for different box sizes and present them in Figure \ref{fig:convergence}.
We plot the BHMF integrated over $z=0-2$ since the GWB amplitude calculation integrates over the SMBHB population in this redshift range.
We also mark where mass bins have fewer than 10 SMBHs with a horizontal dashed line that match the color of the corresponding simulation.
Below the dashed line, the BHMF will be dominated by Poisson noise.
The left plot shows results for IllustrisTNG and MillenniumTNG while the right plot shows results for Simba.
Figure \ref{fig:convergence} also shows the median predicted mass function and the 16th-to-84th percentiles for the CAMELS-TNG and CAMELS-Simba CV sets.
For our purposes, we consider the BHMF of a simulation to be converged if the difference from the largest box simulation (MillenniumTNG and Simba 100 Mpc/h) is within the range expected due to cosmic variance (the 2$\sigma$ range from the CAMELS CV sets).

For the IllustrisTNG and MillenniumTNG simulations, we find that the BH mass functions are converged within 0.5 dex for the simulations with co-moving box lengths of at least 35 Mpc/h at masses $M_\textnormal{BH} > 10^8 M_\odot$. 
At $M_\textnormal{BH} > 10^9 M_\odot$, the variation between these different box size simulations is within the expected range due to cosmic variance, calculated from the CAMELS-TNG CV set, and is therefore considered converged.
Convergence at this mass range is important as higher mass SMBHs will dominate the predicted amplitude.
At masses below $M_\textnormal{BH} = 10^8 M_\odot$ the mass function shows variance of up to a dex between simulations. 
The large variation in the mass function at these masses is likely due to resolution differences between the simulations and will have minimal to no impact on the predicted GWB amplitude. 

The Simba simulations show significant ($<0.1$ dex) agreement between the three different box size simulations up to $M_\textnormal{BH} = 10^{9.5}M_\odot$.
The BHMFs vary more between box sizes at the high mass end ($M_\textnormal{BH} \gtrsim 10^{9.5} M_\odot$) due to under sampling of massive halos in small boxes.
However, the variation between the simulations is within the range expected due to cosmic variance calculated from the CAMELS-Simba CV set and is therefore considered converged.

Although both CAMELS-TNG and CAMELS-Simba fail to produce SMBHs in mass bins larger than $M_\textnormal{BH} \sim 10^{9.5} M_\odot$, the variation between the larger box size simulations is less than the variation expected from the CAMELS CV sets at lower mass bins.
These results imply that a box size of at least (35 Mpc/h)$^3$ for IllustrisTNG or (50 Mpc/h)$^3$ for Simba is required for convergence of the BHMF.

IllustrisTNG and Simba vary dramatically in shape at the low mass ($M_\textnormal{BH} \lesssim 10^8 M_\odot$) end of the BHMF. The IllustrisTNG BHMF rises towards lower masses while the Simba BHMF turns over at these same masses.
This is a result in the difference in how seeding is modeled in the two simulations. 
Since in Simba, BHs are seeded at higher host masses, there will be many less SMBHs at the low mass end than there are in IllustrisTNG which produces the differences seen between their BHMFs.

\begin{figure*}[t!]
    \centering
    \includegraphics[width=\linewidth]{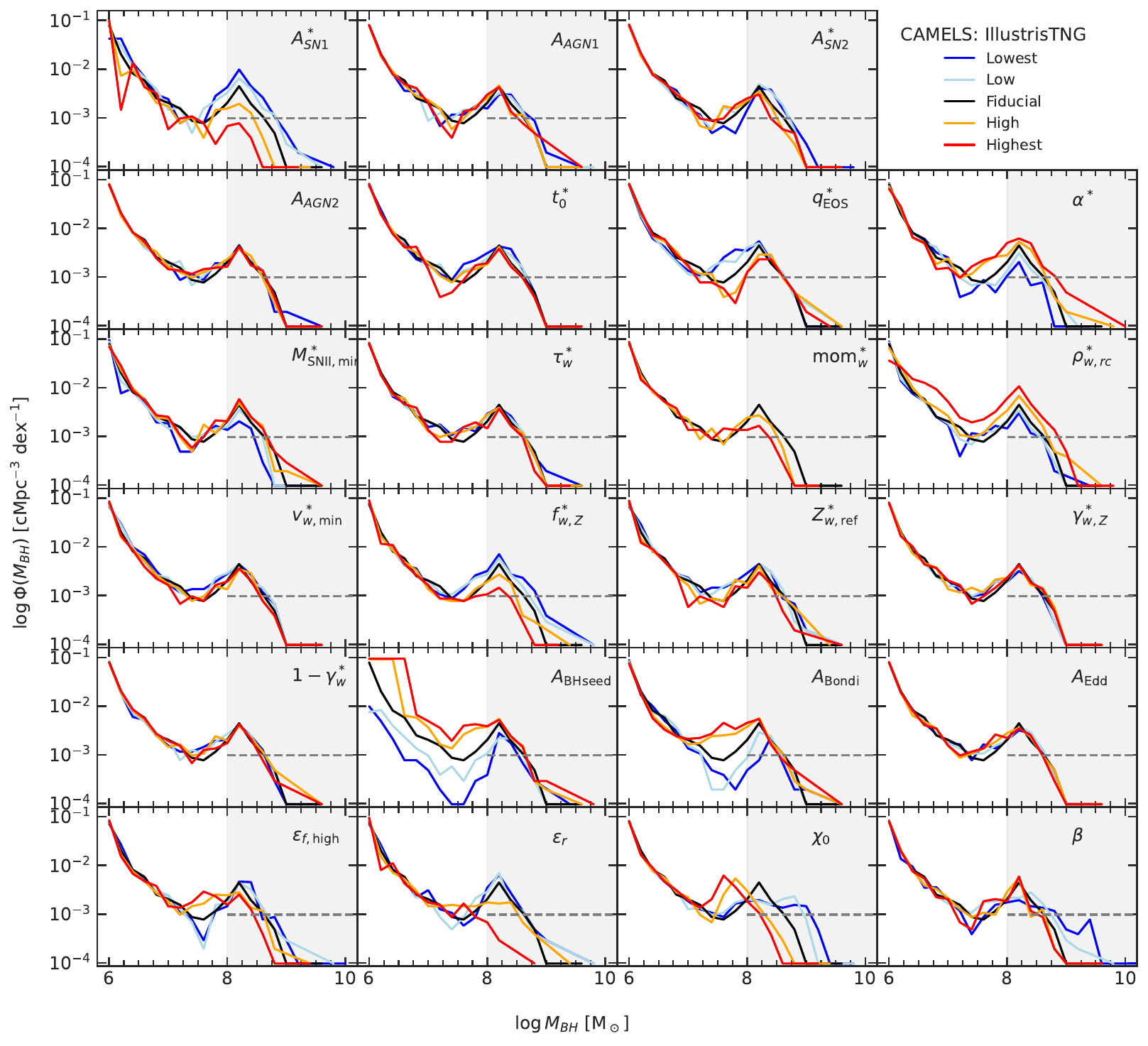}
    \caption{The predicted BHMF at $z=0$ from the CAMELS-TNG simulation 1P set for parameter variations controlling galactic feedback. The name of the parameter varied is labeled in the top right of each plot. The legend indicates what color corresponds to the parameter value with red and blue being the largest and smallest value for that parameter respectively. The grey dashed line marks where the mass bins contain fewer than 10 SMBHs. Some galactic feedback parameters produce large variations in the predicted BHMF while others have minimal effect. Changes to the high mass end of the BHMF are expected to impact the predicted GWB amplitude while changes to the low mass end are not. Descriptions of the parameters can be found in Table \ref{tab:TNGsummary}.}
    \label{fig:TNG_1P}
\end{figure*}

\begin{figure*}[t!]
    \centering
    \includegraphics[width=\linewidth]{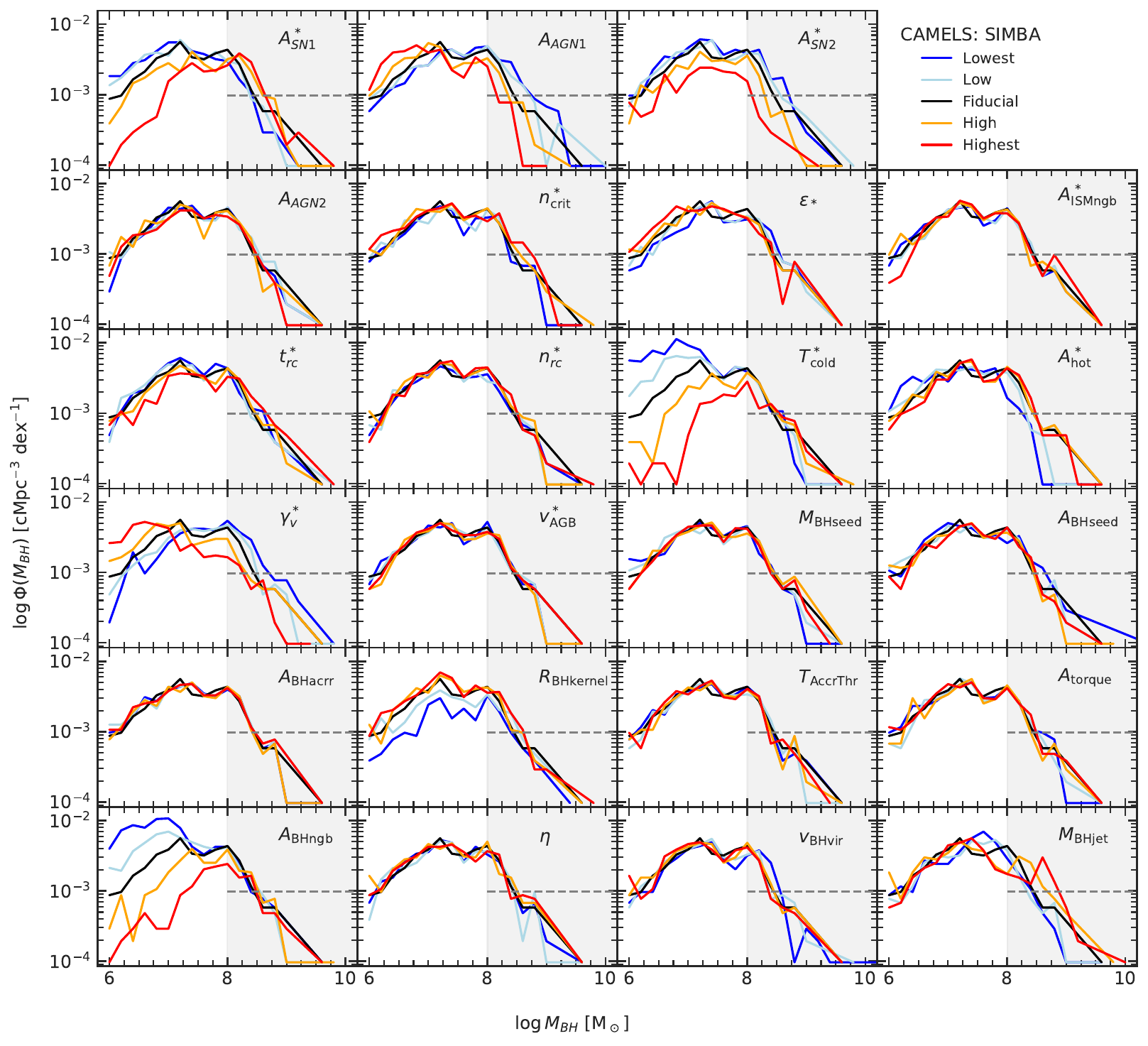}
    \caption{The same as Figure~\ref{fig:TNG_1P} but for the CAMELS-Simba simulation suite. Descriptions of the parameters can be found in Table \ref{tab:SIMBAsummary}.}
    \label{fig:SIMBA_1P}
\end{figure*}

Figures \ref{fig:TNG_1P} and \ref{fig:SIMBA_1P} show the predicted BH mass functions for the CAMELS-TNG and CAMELS-Simba 1P simulation sets respectively.
For each parameter, the mass function is calculated five times for different values of that parameter for which the range of values chosen are discussed in Tables \ref{tab:TNGsummary} and \ref{tab:SIMBAsummary} in the Appendix.
The parameter being varied is labeled in the top right corner in each plot.
Not every parameter variation produces changes in the BHMF that will impact the predicted GWB amplitude.
For example, in Figure \ref{fig:SIMBA_1P}, $A_\textnormal{AGN2}$ has minimal impact on the mass function across the full mass range while WindColdTemp largely impacts the mass function at $M_\textnormal{BH} < 10^8 M_\odot$.
In both cases, the impact on the predicted GWB amplitude will be low.
For both CAMELS-TNG and CAMELS-Simba, many mass functions appear noisy at the highest masses ($M_\textnormal{BH}\sim 10^9 M_\odot$) due to lack of convergence in the small simulation box.

For both CAMELS-TNG and CAMELS-Simba, the high mass end of the BHMF ($M_\textnormal{BH} \geq 10^{8} M_\odot$) is impacted by a mix of stellar and AGN parameters.
For CAMELS-TNG, only $A_\textnormal{BHseed}$ has an impact on the low mass end of the BHMF ($M_\textnormal{BH} < 10^7 M_\odot$).
This is in contrast to CAMELS-Simba where many parameters, both stellar and AGN, impact the low mass end of the BHMF.
Changes to the low mass end of the BHMF will not impact the predicted GWB amplitude but could have implications for other GWB metrics.
We discuss how these different parameters affect the predicted BHMF and thus the GWB amplitude in Section \ref{subsection:discfeedback}.

\subsection{The GWB amplitude}
\subsubsection{The predicted GWB amplitude In Flagship Models}

\begin{figure*}[th!]
    \centering
    \includegraphics[width=0.8\linewidth]{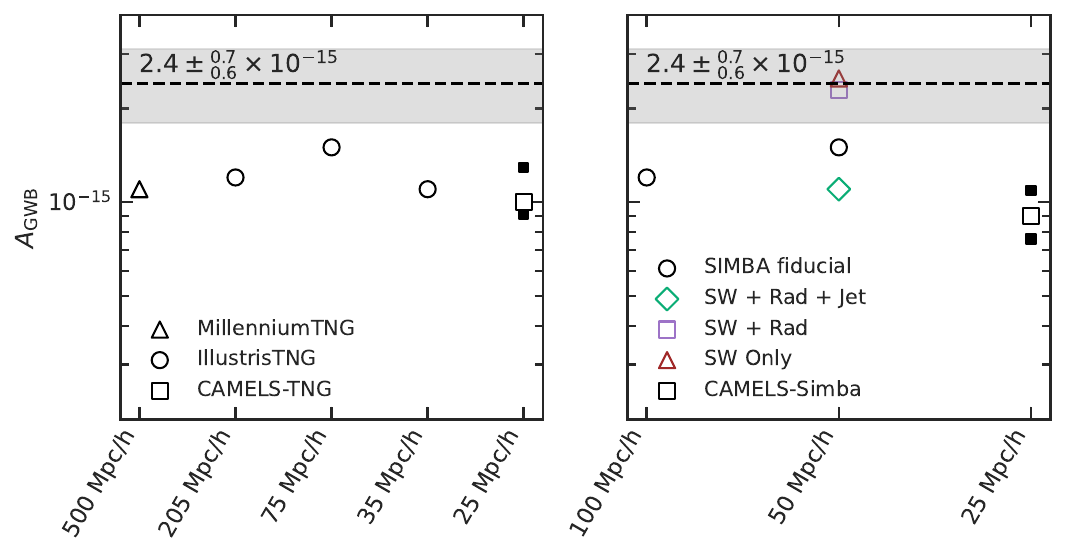}
    \caption{The predicted GWB amplitude for the (left) IllustrisTNG and MillenniumTNG simulations and (right) Simba simulations. The legend denotes which simulation model the predicted amplitude is from and the x-axis denotes the simulation box side-length. The two smaller shaded points for the 25 Mpc/h simulation represents the 16th-to-84th percentiles calculated from the CAMELS CV set. The dashed horizontal line and shaded region corresponds to the PTA-inferred GWB amplitude from the NANOGrav 15 year dataset.}
    \label{fig:GWB_SimbaTNG}
\end{figure*}

Figure~\ref{fig:GWB_SimbaTNG} shows the predicted GWB amplitude for the different box size simulations and the CAMELS CV sets for both IllustrisTNG and Simba.
The figure also shows the predicted GWB amplitude for the feedback variant Simba simulations.
The two smaller points in each figure, at 25 Mpc$/h$, represent the 16th-to-84th percentile ($\sim 2\sigma$) variation predicted by the CAMELS CV set.
Also included is the GWB amplitude value inferred by the NANOGrav 15-yr PTA dataset and a shaded region representing the error \citep{NANOGrav:2023gor}.
The simulation model that the prediction is from is labeled in the legend.

Simulation predicted values of A$_\textnormal{GWB}$ converge beyond 35 Mpc/$h$ for MillenniumTNG/IllustrisTNG and 50 Mpc/$h$ for Simba within the variance of what is expected from cosmic variance (CAMELS 25 Mpc/$h$ CV set simulations).
Importantly all fiducial simulations across a range of volumes (black points) consistently under-predict the GWB amplitude when compared to the measured value from NANOGrav. In our framework, this implies they are missing massive black holes within the $z<2$ redshift range we investigate. 

We note exception in the Simba runs without AGN feedback or certain modes of AGN feedback. 
In particular the Simba SW + Rad and SW only feedback variant simulations, which both correspond to the removal of the AGN feedback jet mode, are consistent with PTA measurements. Recall in the previous section these simulations produced more black holes in the high mass end, explaining this result (see Figure \ref{fig:feedback}). 
In particular, this is a result of the differences seen in the BHMF at $M_\textnormal{BH} > 10^9 M_\odot$, highlighting the significant contribution that the highest BH masses have on the predicted amplitude.

While Simba SW + Rad and SW only feedback variant simulations produce a GWB amplitude consistent with PTA measurements within the \citep{CaseyClyde2022, CaseyClyde2025} framework, they dramatically fail in producing realistic galaxy populations since the jet mode is responsible for quenching \citep{dave:2019,Christiansen:2020}.
The failure of all models in producing both realistic galaxies and a GWB amplitude consistent with PTA measurements points to a need to revise black hole seeding and growth models in cosmological simulations. This points to an agreement between the PTA-inferred GWB amplitude and recent JWST results suggesting that black holes are seeded early and perhaps grew much faster than their host galaxies in the early universe, which is not modeled in these simulations
 \citep[see][and references therein]{2023ApJ...957L...3P}.

\subsubsection{CAMELS Parameter Variation GWB Amplitude}

\begin{figure*}[t!]
    \centering
    \includegraphics[width=\linewidth]{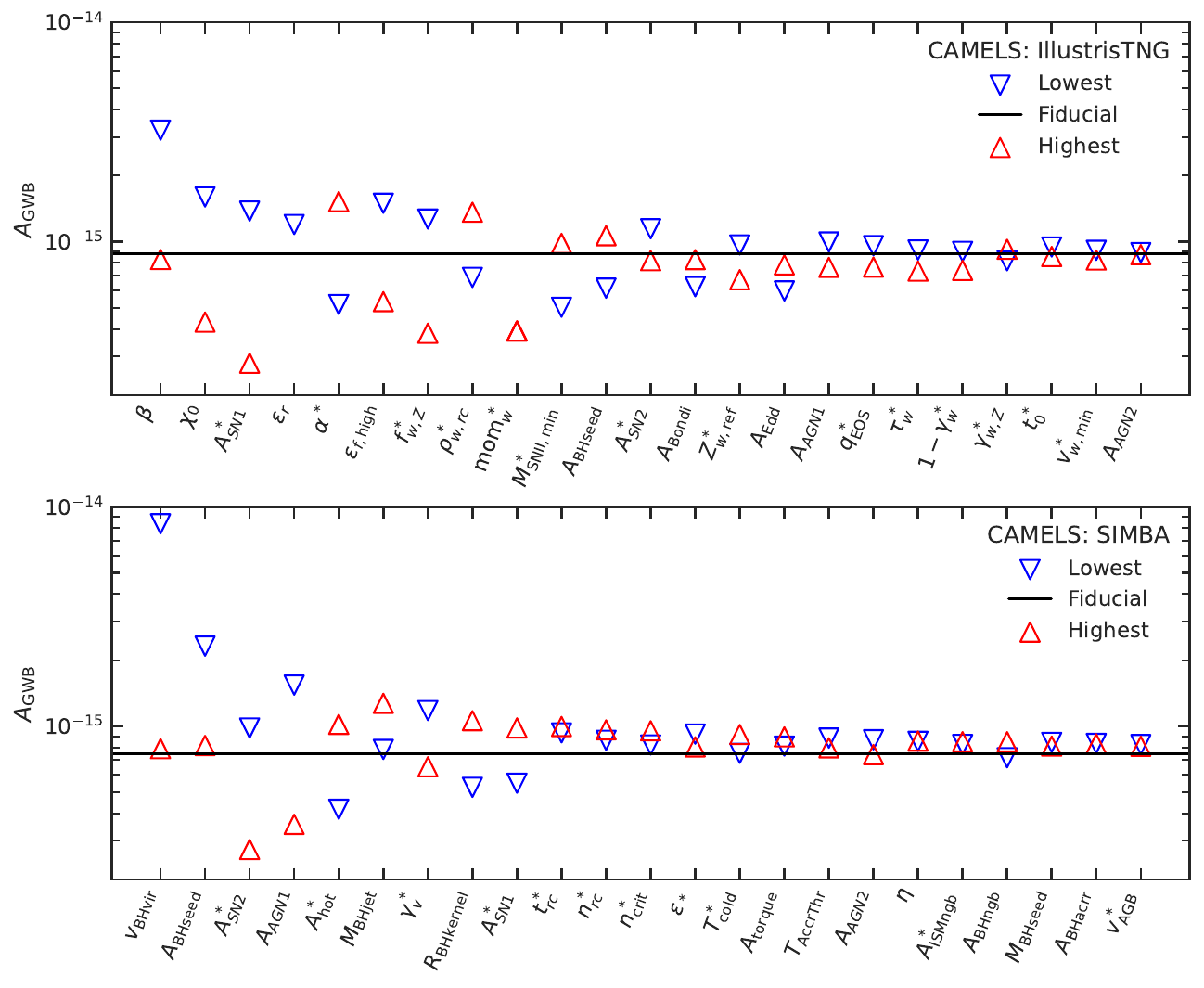}
    \caption{The predicted GWB amplitude from the CAMELS-TNG (top) and CAMELS-Simba (bottom) simulation 1P set for parameter variations controlling galactic feedback. The parameter varied is labeled on the x-axis. Triangles pointing up (red) are for increases in parameter value while triangles pointed down (blue) are for decreases in parameter value. The intermediate parameter values are not plotted but lie between the minimum/maximum and fiducial for all cases. The black solid horizontal line correspond to the fiducial simulation prediction.}
    \label{fig:GWB1P}
\end{figure*}

Figure \ref{fig:GWB1P} shows the predicted GWB amplitude values from the CAMELS-TNG and CAMELS-Simba 1P sets. Descriptions of these parameter variations can be found in Tables 1 and 2. 
We order the parameters in these plots from the highest impact on the predicted GWB amplitude on the left side  to the lowest impact towards the right side.
For each simulation, only one parameter produces a large enough change in the predicted GWB amplitude to reproduce the measured value, $\beta$ for CAMELS-TNG and $A_\textnormal{BHseed}$ for CAMELS-Simba. 
However, we generally refrain from comparing these predictions to the PTA-inferred GWB directly in the plot for two reasons: 
One is the lack of convergence in the CAMELS simulations due to the small box size.
Second is that variations due to the highest and lowest parameter values in the CAMELS 1P set often produce non-physical results such as a mismatch between the observed and predicted stellar mass function.
Due to these facts, this analysis serves largely to analyze the extent to which galactic feedback may impact the predicted GWB amplitude.
These results also emphasize that even when varying these parameters to extreme values, these simulations still struggle to reproduce the measured amplitude.
For a robust comparison with observations, the simulations must be run in a larger box and furthermore confirmed to reproduce realistic galaxies given the different parameter variations.
For this work, we include a brief discussion on how physically realistic the galaxy populations are for the two CAMELS simulations that get closest to the PTA prediction in Section \ref{ss:obsdis}.

\begin{figure*}[t!]
    \centering
    \includegraphics[width=0.7\linewidth]{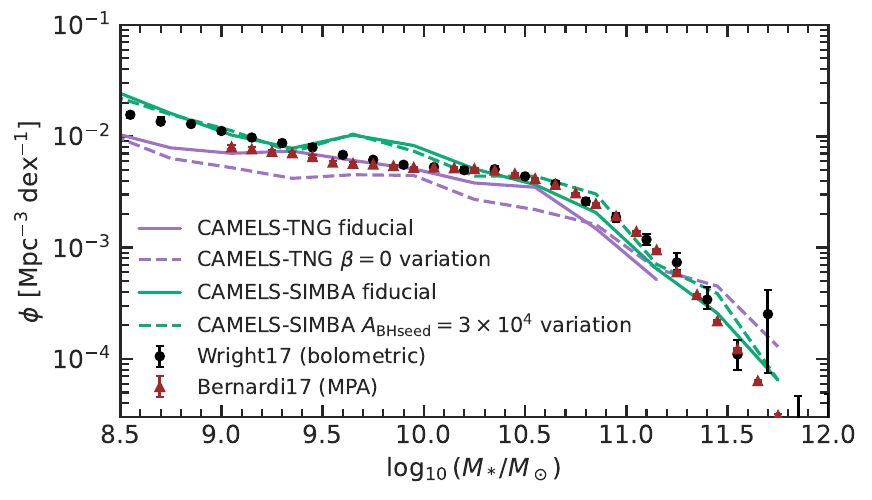}
    \caption{The predicted stellar mass function from the two CAMELS simulation parameter variations that best match the recent NANOGrav measurements (CAMELS-TNG $\beta = 0$ and CAMELS-Simba $A_\textnormal{BHseed} = 3\times10^4$). The solid lines are the fiducial simulation predictions and the dashed lines are the parameter variations both labeled in the legend. The simulated stellar mass functions are compared to the observed stellar mass functions (scatter points) from \citet{Bernardi:2017} and \citet{Wright:2017}.}
    \label{fig:SMF}
\end{figure*}

\section{Discussion}
\label{sec:discussion}

\subsection{Implications for Massive Black Hole Growth}

The fiducial IllustrisTNG, MillenniumTNG, and Simba simulation models all under-predict the observed GWB amplitude using the quasar framework. 
This comes from a lack of SMBHs populating the highest mass range ($M_\textnormal{BH} \geq 10^8 M_\odot$) as seen through the BHMF. 
Some simulations with varying feedback prescriptions predict a GWB amplitude that matches what is measured via PTAs.
However, those variations are unrealistic as they do not include or suppress mechanisms that regulate SMBH growth and as a result those simulations do not reproduce other galaxy properties. 
Instead, we can examine those simulations to determine the number of SMBHs that would be required by the fiducial models to resolve simulation predictions and PTA measurement of the GWB amplitude.
The CAMELS 1P set simulations are not converged at the highest mass ranges of the BHMF so we conduct this exercise with the Simba feedback variant simulations which are converged with a box length of 50 Mpc/$h$.
The SW only and the SW + Rad simulations match the PTA measurements. 

We will look specifically at the SW + Rad simulation as it is more realistic including both stellar feedback and a form of AGN feedback.
Looking at how many more SMBHs this simulation has in different mass bins when compared to the fiducial $50$ Mpc/$h$ Simba box will demonstrate to what degree the most massive SMBHs are under-predicted in the fiducial Simba model.
For the mass bin of $10^8< M_\textnormal{BH}<10^{8.5} M_\odot$, the SW + Rad simulation has $\sim 1.5$ times as many SMBHs as the fiducial case.
For $10^{8.5}< M_\textnormal{BH}<10^{9} M_\odot$ the SW + Rad simulations has $\sim 2.5$ times as many SMBHs, for $10^{9}< M_\textnormal{BH}<10^{9.5} M_\odot$ there are $\sim 5.5$ times as many SMBHs, and for $M_\textnormal{BH}>10^{9.5} M_\odot$ there are $\sim 3.5$ times as many SMBHs.

Overall, the fiducial Simba simulation requires $\sim1.5$ to $5.5$ times more of the most massive SMBHs ($M_\textnormal{BH} > 10^8 M_\odot$), depending on the mass bin, to reproduce PTA measurements.
Since the higher mass BHs contribute more to the GWB amplitude, a similar amplitude can be predicted with, for example, less $M_\textnormal{BH}>10^{9.5} M_\odot$ BHs if there are even more $M_\textnormal{BH}<10^{9.5} M_\odot$ BHs. 
However, those lower mass BHs are less impactful in their contribution to the GWB, so many more are required.
When utilizing the local SMBH population to calculate the GWB amplitude, \citet{Sato-Polito:2023, Sato-Polito:2025} similarly found that more SMBHs were necessary, requiring up to 10 times more SMBHs than local measurements suggests to reproduce the measured amplitude SMBHs.
They found that SMBHs with masses $M_\textnormal{BH} \gtrsim 10^{10} M_\odot$ contributed the most, but that current amplitude measurements disfavors a BHMF that is dominated by a few very heavy sources. 
Furthermore, \citet{Mingarelli2026} found that one requires these very massive $10^{10}~M_\odot$ SMBH binaries to reproduce the PTA-inferred GWB, using conservation of energy arguments.

These results highlight the need to re-evaluate current models for SMBH growth and seeding to determine how to grow these missing massive SMBHs in state-of-the-art cosmological hydrodynamic simulations.
\citet{Liepold&Ma:2024} recently derived the BHMF from the galaxy stellar mass function using updated observations find a higher amplitude of the most massive stellar mass galaxies and thus SMBHs.
The inferred abundance of SMBHs with masses above $10^{10}M_\odot$ was found to be consistent with both the number of currently known systems and the predicted amplitude of the stochastic GWB.
Simulations often calibrate their models to the observed stellar mass function, therefore a model calibrated to the mass function found by \citet{Liepold&Ma:2024} would likely predict an amplitude closer to PTA measurements. 
However, one caveat is that the \citet{Liepold&Ma:2024} BHMF yields a SMBH mass density at $z=0$ that is inconsistent with the value estimated from quasars at higher redshifts.

We emphasize that missing massive SMBHs is not the only possible explanation for the deficit. The GWB amplitude is degenerate with the binary merger rate; a higher binary duty cycle than inferred by our $\mathcal{N}_{cal}$ calibration could similarly boost the signal. Additionally, deviations from the assumed $-2/3$ spectral index could alter the extrapolated amplitude at $f=1\,\mathrm{yr}^{-1}$. 
Finally, \citet{Lapi:2026} argue, with a semi-empirical approach utilizing both the local BHMF and galaxy pair counts from JWST, that 50-70\% of the stochastic GWB signature could be accounted for in sources other than SMBHBs, either astrophysical or cosmological.

\subsection{Stellar and AGN feedback and the GWB Amplitude}\label{subsection:discfeedback}

The link between the GWB amplitude, the binary massive black hole population and, in turn, the massive black hole population provides an important test for galaxy formation models, since the high-mass BHMF is shaped by feedback physics that governs black hole growth and additionally to the black hole seeding prescription at high redshift.  

Our results demonstrate that the choice of feedback model (including both AGN and stellar/SNe feedback) has a strong impact on the BHMF in cosmological box simulations and hence the predicted GWB amplitude using the quasar framework.  
In simulations with inefficient AGN or stellar feedback, SMBHs can grow rapidly in massive halos, yielding an overabundance of extremely massive black holes (e.g., $M_\textnormal{BH}\gtrsim10^9$–$10^{10}M_\odot$). 
Such models predict a significantly higher GWB amplitude than models with strong feedback. 
In contrast, when energetic AGN and/or SNe feedback is included, it self-regulates black hole accretion in large halos and curtails the growth of the most massive black holes. 
This produces a cut-off or flattening in the BHMF at the high-mass end, which in turn lowers the GWB amplitude.

Our findings on the BHMF are consistent with previous studies on the behavior of BH growth in cosmological simulations. 
Large-volume simulations of galaxy formation, such as Illustris, EAGLE, IllustrisTNG, and Simba, all require some form of AGN feedback to reproduce galaxy and star formation properties, but they differ in the details of BH growth outcomes. 
For example, the original Illustris simulation (which employed thermal bubble AGN feedback) yielded very massive BHs in group and cluster-scale halos, overshooting some observational constraints on BH masses \citep{Sijacki2015}. 
The IllustrisTNG model introduced a kinetic AGN feedback mode that more efficiently quenches massive galaxies, thereby slowing down BH growth at late times \citep{Weinberger2018, Pillepich:2018a}. As a result, IllustrisTNG produces fewer ultra-massive black holes at $z=0$ than Illustris.
The EAGLE simulation, which calibrates AGN heating to reproduce the $M_\textnormal{BH}$–$\sigma$ relation, similarly finds that strong AGN feedback is needed to prevent excessive BH buildup in the most massive galaxies \citep{Schaye2015, Rosas-Guevara2015, McAlpine2018}. 
By contrast, if AGN feedback is turned off or greatly weakened (leaving only stellar feedback), simulations tend to produce overly massive central BHs. 

Using the Simba simulation as an illustrative case: our tests show that removing kinetic AGN jet feedback leads to a factor of a few increase in the number density of $M_\textnormal{BH}>10^8,M_\odot$ black holes (relative to the fiducial Simba model with jets), and completely turning off all AGN feedback causes an even more dramatic upturn in the high-mass BHMF. 
Such variations in the high-mass tail of the BHMF translate almost directly into GWB amplitude differences. 
The right plot of Figure~\ref{fig:GWB_SimbaTNG} of our results (see above) explicitly demonstrated that models with no AGN or reduced feedback yield a substantially higher background strain than those with efficient AGN feedback, owing to the surplus of massive BH binaries in the former case. 

We see this demonstrated in Figure~\ref{fig:GWB1P} with the most impactful parameters for the GWB amplitude being $\beta$ and $\chi_0$ for IllustrisTNG and $v_\textnormal{BHvir}$ for Simba.
These parameters directly control the strength of the different AGN feedback models implemented in the simulations.
This demonstrates the capabilities for AGN feedback to create a self-regulating effect on SMBH growth.
The next most impactful parameter on the GWB amplitude for Simba is $A_\textnormal{BHseed}$, which more directly impacts the BHMF by controlling the mass at which SMBHs are seeded.
While some of these parameter variations predict a higher background strain that better matches the measured value from PTAs, the extreme parameter variations in the CAMELS simulations are often ruled out through their inability to reproduce various observed galaxy statistics.
We discuss this further in section \ref{ss:obsdis} with a particular focus on the two parameter variations that best match the PTA measurements: $\beta = 0$ for CAMELS-TNG and $A_\textnormal{BHseed} = 3\times10^4$ for CAMELS-Simba. 

It is worth noting that while many stellar feedback parameters primarily affect the low-mass end of the BHMF—for instance, stronger SN-driven winds can delay BH seed growth in galaxies \citep{Dubois2015,AnglesAlcazar2017,Ravi2019}, our analysis shows that some stellar feedback parameters can, in fact, influence the predicted gravitational wave background.  
In particular, in the Simba suite we find that the normalization factor for the speed of galactic winds ($A_\textnormal{SN2}^*$) and the fraction of wind particles ejected in the hot phase ($A^*_\textnormal{hot}$) significantly affect the resulting GWB amplitude. 
In the IllustrisTNG simulations, the most important stellar feedback parameters are the normalization factor for the energy in galactic winds per unit star formation ($A^*_\textnormal{SN1}$) and the power-law index of the stellar initial mass function above 1 $M_\odot$ ($\alpha^*$). 
For both Simba and IllustrisTNG parameters affect how efficiently stellar-driven outflows regulate black hole growth, and thus indirectly shape the abundance of SMBHs in the PTA-sensitive mass range. Moreover, our framework relies on quasars as tracers for the binary population. Consequently, AGN feedback plays a dual role in our predictions: it physically limits the black hole mass (shaping the BHMF) and, by suppressing accretion rates, effectively reduces the number of visible tracers used to infer the binary population. This double suppression implies that strong feedback models may underestimate the GWB if a significant population of dark binaries exists in quenched galaxies.

\subsection{Comparisons to Observations}\label{ss:obsdis}

As discussed above, there are two CAMELS simulation variations that predict a GWB amplitude close to that inferred by PTA measurements. 
These two simulations are: the CAMELS-TNG simulation that sets QuasarThresholdPower to the lowest value ($\beta = 0$) and the CAMELS-Simba simulation that sets the BHSeedRatio to the lowest value ($A_\textnormal{BHseed} = 3\times10^4$).
While these simulations are consistent with PTA measurements, we must consider if they still reproduce realistic galaxy populations.


For the IllustrisTNG model, $\beta$ controls the switch between the low- to high-accretion state and thus which AGN feedback mode is active at what times. In the case of $\beta = 0$, lower mass SMBHs are more likely to be producing kinetic feedback, the mode primarily responsible for quenching star formation, and higher mass SMBHs are more likely to be in QSO mode, the mode where SMBHs grow most efficiently.
Allowing higher mass SMBHs to grow more efficiently explains why this variation matches PTA measurements better, given that higher mass binaries are necessary to increase the predicted GWB amplitude. 
This mirrors a result found by \citet{Barausse:2023} where SMBHs are required to accrete more efficiently to reproduce PTA measurements.
However, activating the kinetic feedback mode more often in lower mass SMBHs, and thus lower mass galaxies, will risk quenching galaxies before they reach sufficiently high stellar masses when comparing to observations.
This earlier suppression of stellar feedback could also contribute to the ability of SMBHs to efficiently grow without their fuel source begin disrupted.

For the Simba model, decreasing $A_\textnormal{BHseed}$ to $3\times10^4$ means that SMBHs are being seeded in galaxies with stellar masses an order of magnitude lower than the fiducial model.
SMBHs are still seeded at a seed mass of $10^4 M_\odot/h$ which is not necessarily unrealistic for galaxies with stellar masses $\sim 10^8 M_\odot/h$.
Earlier seeding allows SMBHs to grow for longer times, enabling them to reach the higher masses required to reproduce the PTA measurements.
This resembles a result found by \citet{Bonoli:2025} where they found that lighter seeds ($\sim 100 M_\odot$) that grow rapidly and efficiently in the early universe were required to reproduce PTA measurements. 
A potential conflict in this variation comes from the AGN feedback model now operating in lower mass galaxies which may cause a disruption to early star formation that may impact the resulting galaxy population.

We examine the stellar mass function (SMF) predicted by these simulations to consider how physically realistic the galaxies produced with these parameter variations are. 
Figure \ref{fig:SMF} plots the predicted SMF from the fiducial CAMELS simulations and the parameter variant CAMELS simulations that best match the PTA-inferred GWB.
We compare these predictions to the observed SMF from \citet{Bernardi:2017} and \citet{Wright:2017}.
The CAMELS-TNG $\beta$ parameter variation struggles to match the observed SMF, under-predicting the low mass end. 
This implies that the change made to the AGN feedback model through varying $\beta$ did quench low mass galaxies to a degree that makes that variation unrealistic.

By contrast, the CAMELS-Simba $A_\textnormal{BHseed}$ variation produces a SMF that largely matches that of the fiducial simulation.
While changing the way that SMBHs are seeded in the Simba model will impact how AGN feedback is introduced, in this case it does not appear to have a strong enough effect to impact a global statistic such as the SMF.
Figure 17 of \citet{Ni:2023} provides additional insight on this parameter variation for a different statistic, the gas power spectrum.
The minimum value for $A_\textnormal{BHseed}$ results in a gas power spectrum with more power on larger physical scales and less power on small scales when compared to the fiducial prediction.
This implies that introducing SMBH seeds to lower mass galaxies causes a redistribution of gas from the center of halos to larger scales.
This could be due to the introduction of AGN feedback in weaker potential wells from earlier SMBH seeding.
While the predicted SMF still matches observations, other observables will likely be impacted by this change evident by the change seen in the gas power spectrum.

To summarize, while changes to both SMBH growth/seeding models and galactic feedback models can impact the predicted GWB amplitude, these changes also affect other observables, often in negative ways. 
We compared these parameter variations using the SMF because it is a statistic that cosmological hydrodynamic simulations typically use to constrain their models, but many more observables are likely disrupted by these variations. 
For example, SMBH statistics such as the local occupation fraction, the quasar luminosity function, and the local BHMF are all potentially impacted, and testing these would require a comprehensive study involving assumptions about mechanisms such as the SMBH duty cycle that are not typically explicitly modeled in these simulations. 
Indeed, previous work found that increasing SMBH accretion to match the PTA-inferred GWB struggles to simultaneously reproduce the observed quasar luminosity function and local BHMF \citep{Izquierdo-Villalba:2022}. 
Because galactic feedback models are explicitly designed to regulate star formation and match observed galaxy statistics, changes directly to these models are the most likely to produce unrealistic galaxy populations. 
Changes made to SMBH seeding or growth prescriptions, while still intrinsically tied to AGN feedback, carry a degree of separation from the feedback mechanisms themselves, making such variations somewhat more likely to reproduce realistic galaxy populations across a broader range of observables.

\subsection{Assumptions in modeling}

Many studies use simulation merger trees directly to predict the GWB \citep[e.g.,][]{Kelley2017b, Sykes2022, Chen2025, Marinacci:2025}. We instead adopt the quasar-based framework of \citet{CaseyClyde2022} since we are asking how AGN-feedback reshapes the BHMF, and how this propagates into the GWB amplitude. The BHMF is the only simulation input in this framework. Here, merger timescales and SMBH pairing rates are absorbed into a fixed empirical calibration, $\mathcal{N}_{\rm cal}$). This is particularly important for the $25~{\rm Mpc}/h$ CAMELS boxes, where merger tree statistics at the high-mass end are too sparse for a direct merger rate calculations. Merger tree predictions also depend on sub-grid SMBH dynamics (e.g. dynamical friction, stellar hardening, circumbinary torques) that vary between simulations and are complementary to the BHMF physics we focus on here. However, we do inherit the assumptions of \citet{CaseyClyde2022, CaseyClyde2025}, and we discuss the impact of those simplifying assumptions required to map the simulated BHMF to a predicted GWB amplitude below. A direct comparison of both methods is the subject of a follow-up study.

First, we assume that the instantaneous BHMF from the simulations directly represents the population of SMBHs participating in binaries that contribute to the nanohertz GWB. 
This implicitly neglects details of merger timing, binary pairing efficiency, and environmental hardening processes that depend on each galaxy’s dynamical history.
Since binary evolution occurs on Gyr timescales, the GWB amplitude is dominated by systems assumed at higher redshifts.
However, the impact of time delays on the amplitude is complex. While long delays can prevent binaries from merging by $z=0$, thereby suppressing the signal, they also shift the merger epoch of high-redshift binaries to lower redshifts. Since the GW strain scales as $1/d_L$, binaries merging at lower redshift have a larger strain. Therefore, accounting for time delays could either suppress or enhance the predicted amplitude by shifting mergers to the local Universe, depending on the specific delay time distribution.

Second, we assume that the empirical relation between quasar luminosity and BH mass can be used to weight the simulated BHMF via the quasar luminosity function, effectively treating quasar activity as a tracer of binary incidence. This step assumes that all quasars of a given mass have equal likelihood of forming binaries, and that the redshift evolution of the quasar population mirrors that of SMBHs.

Third, we assume the empirical duty cycle to be constant across all models which neglects how the physical duty cycle in simulations, including the extreme feedback variations, diverge from that calibration.
This mapping assumes that quasar visibility remains constant across all models.

Fourth, the model prescribes a lognormal mass-ratio distribution and circular binary orbits, omitting eccentricity evolution and spin coupling, both of which can alter the spectral shape and amplitude of the GWB.

Future work should compare our method here to those that couple simulation-derived BHMFs with self-consistent merger trees that explicitly track SMBH pairing, dynamical friction, and binary hardening within evolving galaxies.
Incorporating such merger histories would allow us to directly compute the SMBHB formation rate and residence time at PTA frequencies, providing a more physically motivated connection between simulated BH growth and the PTA-inferred GWB amplitude \citep[e.g.,][]{Sesana2013,Kelley2017b,Barausse2020,Chen2025}.
Previous work analyzing the Illustris simulation took care to include many of the environmental mechanisms expected to impact GWB predictions \citep[dynamical friction, stellar `loss-cone' scattering, viscous drag from a circumbinary disc, and eccentric binary evolution][]{Kelley2017a,Kelley2017b} and found an amplitude of up to $10^{-15}$ at a frequency of 1 yr$^{-1}$.

Even when considering these additional factors, despite the method use, all predictions reported from cosmological simulations under-predict the value reported by the NANOGrav 15-year dataset.
Considering these additional enviornmental factors also seem to reduce the predicted amplitude rather than increase it, at least for Illustris.
When calculating the amplitude predicted by Illustris with our model we find an amplitude of $1.6\times 10^{-15}$, which is higher than that found by \citet{Kelley2017b}.
This is due to the fact that at the massive end of the BHMF, Illustris has more massive SMBHs than IllustrisTNG in the relevant redshift range.
This comparison implies that it is important to consider these environmental mechanisms when calculating the GWB amplitude, but that the consideration of these mechanisms alone cannot resolve the tension between simulation predictions and PTA measurements.

Finally, works following this one should aim to establish a self-consistent framework for predicting the GWB amplitude from simulations which considers these environmental mechanisms and utilizes merger histories.
A recent study built a framework for building gravitational wave event catalogs by assigning merger events to stellar particles in a simulation and applied it to MillenniumTNG \citep{Marinacci:2025}.
Follow-up studies could utilize such a tool for a more self-consistent analysis of the GWB in cosmological hydrodynamic simulations.

\section{Conclusions}
\label{sec:conclusion}

We have developed a framework that connects simulation-predicted SMBH mass functions to the GWB amplitude measured by PTAs. 
Building on the quasar-based population model of \citet{CaseyClyde2022}, we derive an estimate of the SMBHB population using the subset of SMBHs that would appear as active quasars. \citet{CaseyClyde2025} found that a quarter of these will be in binary form.  This then allows us to compute the GWB amplitude.  
We apply this framework to the IllustrisTNG, MillenniumTNG, Simba and CAMELS simulation suites spanning volumes from $25$ to $500\ \mathrm{Mpc}/h$ and covering a range of astrophysical models.
We emphasize that while recent works have revealed that PTA measurements can act as a probe of SMBH populations and thus provide constraints on SMBH seeding and growth models, the role that galactic feedback has on influencing the SMBH population cannot be ignored. This work in particular shows that PTA measurements can also indirectly constrain galactic feedback models.

Our main conclusions are as follows:

\begin{itemize}

    \item We predict a value for the GWB amplitude ($A_\textnormal{GWB}$) at a reference frequency of 1 yr$^{-1}$ in various cosmological simulations. For IllustrisTNG and MillenniumTNG, $A_\textnormal{GWB}$ spans $1.1 - 1.5\times 10^{-15}$ for 35-500 Mpc/$h$ boxes.  For Simba $A_\textnormal{GWB}$ spans $1.2 - 1.5\times 10^{-15}$ for 50-100 Mpc/$h$ boxes.
    \item We find convergence in the predicted GWB amplitude for volumes larger than $\sim(35-50$ Mpc$/h)^3$. At $(25$ Mpc$/h)^3$ we find that cosmic variance results in a 2$\sigma$ deviation of $\sim 0.1$ dex in the predicted GWB amplitude and significant noise in the BHMF for $M_\textnormal{BH}>10^9 M_\odot$, a mass range vitally important for predictions of the amplitude. 
    \item The IllustrisTNG, MillenniumTNG, and Simba fiducial simulation models all under-predict the GWB strain amplitude by a factor of $\sim 2$ and require 5.5 times more of the most massive SMBHs ($M_\textnormal{BH} > 10^9 M_\odot$) to reproduce the measured strain. 
    \item Varying the underlying AGN and stellar feedback prescriptions produces systematic differences in both the shape and normalization of the BHMF. These differences can impact the predicted GWB strain amplitude by a factor of 2 for the Simba feedback variant simulations and up to a factor of 10 for CAMELS extreme feedback variations.
    \item The feedback parameters most critical to the high-mass end of the BHMF, and hence to the GWB amplitude, include, but are not limited to, the normalization of AGN feedback energy, the speed and thermal fraction of stellar winds, and the high-mass slope of the stellar initial mass function. These parameters regulate SMBH growth in massive galaxies and thus control the abundance of binaries that dominate the PTA frequency band.
\end{itemize}

Comparison with PTA measurements illuminates a new empirical constraint on SMBH growth and feedback physics in large scale simulations. 
Matching the GWB amplitude measured by PTAs requires simulations to produce realistic high-mass SMBH abundances consistent with both electromagnetic and GW measurements.
Since multiple galactic feedback models can reproduce realistic galaxy populations, introducing novel metrics for testing galactic feedback models, such as PTA measurements, will be necessary to further constrain our understanding of galactic feedback and its role in shaping the Universe.
Two Simba feedback variant simulations and two CAMELS simulations were able to reproduce the expected signal by removing or heavily suppressing AGN and stellar feedback models.
However, these simulations do not reproduce observed galactic statistics and are thus ruled out as physically realistic models.
Two CAMELS parameter variation simulations were able to reproduce PTA measurements.
The CAMELS-TNG simulation that changed the scaling of the low- to high-accretion state threshold for SMBHs struggled to reproduce the observed SMF.
The CAMELS-Simba simulation that changed the stellar mass at which SMBHs are seeded reproduced the observed SMF but could still struggle to reproduce other observables.
Therefore, SMBH growth/seeding and galactic feedback models in cosmological hydrodynamic simulations need to be re-evaluated in order to determine how to effectively reproduce the necessary SMBH population required to reproduce PTA measurements.

While this work reported the GWB amplitude at a reference frequency of 1 yr$^{-1}$, analyzing the amplitude at different frequencies or over the full characteristic strain spectrum could provide further insight into the impact that galactic feedback has on the GWB. 
This type of analysis would benefit more from the inclusion of environmental factors, such as binary hardening, merger timing, and pairing efficiency, as many of these processes are expected to impact the highest and lowest frequencies of the characteristic strain spectrum. 
Another factor to consider is the cosmological parameters implemented in these simulations.
While we did not explore the impact of these assumptions, the CAMELS simulations do explore some variation in cosmological parameters such as $\sigma_8$ and $\Omega_m$.
Changes to dark matter or dark energy models could affect the growth of density perturbations which could propagate to affect early halo growth and SMBH seeding.
Earlier SMBH seeding could give these objects more time to grow and relax the need for more extreme accretion scenarios.

This study explores the empirical bridge between cosmological simulations, quasar demographics, and the GWB. 
Our framework assumes that the instantaneous BHMF reflects the population of binary SMBHs and that quasar activity traces SMBHB incidence. 
These simplifying assumptions neglect merger timing, pairing efficiency, and binary hardening processes, which future work should incorporate using self-consistent merger tree modeling. 
Coupling simulation-based BHMFs with detailed binary evolution physics will enable a more predictive description of the GW Universe.

\begin{acknowledgments}
B.B. acknowledges support from NASA grant 80NSSC24K0773  (ATP23-0149) and NASA grant No. 80NSSC20K0500. B.B. acknowledges NSF grant AST-2407877.
B.B. is grateful for the generous support by the David and Lucile Packard Foundation and the Alfred P. Sloan Foundation. 
B.B. and C.M.F.M. thank the Center for Computational Astrophysics (CCA) of the Flatiron Institute and the Mathematics and Physical Sciences (MPS) division of the Simons Foundation for support.  The Flatiron Institute is supported by the Simons Foundation.
The authors thank Laura Blecha for valuable comments. C.M.F.M.\ was supported in part by the National Science Foundation under Grants No.\ NSF PHY-1748958, AST-2106552, and NASA LPS 80NSSC24K0440. 
S.B. is supported by the UKRI Future Leaders Fellowship [grant numbers MR/V023381/1 and UKRI2044]
E.R. is supported by NKFIH-OTKA grant K-142534 from the National Research, Development and Innovation Office (NKFIH), Hungary.
\end{acknowledgments}

\appendix

Presented here are the names, symbols, and descriptions of the CAMELS 1P simulation set parameters explored in this work. 

\nolinenumbers

\begin{table*}\caption{Summary of the galactic feedback parameters in the CAMELS-IllustrisTNG suite. Values are sampled logarithmically unless otherwise stated. Parameters with an asterisk correspond stellar modeling while those without correspond to SMBH modeling. More details about the parameters can be found in \citet{Ni:2023}.}\label{tab:TNGsummary}
\centering
\begin{tabularx}{\textwidth}{l|c|P{0.6\textwidth}}
\hline
\multicolumn{1}{c}{Name from Ni et al. 2023} & \multicolumn{1}{c}{Symbol} & \multicolumn{1}{c}{Description}  \\ \hline \hline
WindEnergyIn1e51erg                          & $A_{SN1}^*$                        & Normalization factor for energy in galactic winds per unit SF. Sampled between 0.9 and 14.4 with fiducial value 3.6.  \\ \hline
VariableWindVelFactor                        & $A_{SN2}^*$                        & Normalization factor for the galactic wind speed. Sampled between 3.7 and 14.8 with fiducial value 7.4.  \\ \hline
RadioFeedbackFactor                          & $A_{AGN1}$                       & Normalization factor for the energy in AGN feedback per unit accretion rate in the low-accretion state. Sampled between 0.25 and 4 with fiducial value 1.   \\ \hline
RadioFeedbackReorientationFactor            & $A_{AGN2}$                       & Normalization factor for the frequency of AGN feedback energy release events in the low-accretion state. Sampled between 10 and 40 with fiducial value 20.                                                            \\ \hline
MaxSfrTimescale                              & $t^*_0$                          & Timescale for SF at the density threshold of SF. Sampled between 1.135 and 4.54 Gyr with fiducial value 2.27 Gyr                                                                                                      \\ \hline
FactorForSofterEQS                           & $q_\textnormal{EOS}^*$                        & Interpolation factor between the effective equation of state for SF gas. Sampled between 0.1 and 0.9 with fiducial value 0.3.                                                                                         \\ \hline
IMFslope                                     & $\alpha^*$                         & Power-law index of the stellar initial mass function above 1\$M\_\textbackslash{}odot\$. Sampled linearly between -2.8 and -1.8 with fiducial value -2.3.                                                                             \\ \hline
SNII\_MinMass\_Msun                          & $M_\textnormal{SNII,min}^*$        & Lower threshold for the mass of a star that produces a SN explosion. Sampled linearly between 4 and 12 $M_\odot$ with fiducial value 8 $M_\odot$.                                                                                     \\ \hline
ThermalWindFraction                          & $\tau_w^*$                         & Fraction of galactic wind feedback energy that is injected thermally. Sampled between 0.025 and 0.4 with fiducial value 0.1.                                                                                          \\ \hline
VariableWindSpecMomentum                     & $\textnormal{mom}_w^*$             & Normalization factor for the specific momentum in galactic winds per unit SF. Sampled linearly between 0 and 4000 km s$^{-1} $ with fiducial value 0 km s$^{-1}$.                                                                      \\ \hline
WindFreeTravelDensFac                        & $\rho_{w,rc}^*$                    & Sets the gas density around (collisionless) galactic wind particles at which they recouple back into the hydrodynamics. Sampled between 0.005 and 0.5 with fiducial value 0.05 in units of density threshold for SF.  \\ \hline
MinWindVel                                   & $v_{w,\textnormal{min}}^*$         & Minimum value imposed for the galactic wind speed. Sampled linearly between 150 and 550~km s$^{-1}$ with fiducial value of 350~km s$^{-1}$.                                                                                           \\ \hline
WindEnergyReductionFactor                    & $f_{w,Z}^*$                        & Normalization factor for the energy of galactic winds at high compared to low metallicity. Sampled between 0.0625 and 1 with fiducial value 0.25.                                                                     \\ \hline
WindEnergyReductionMetallicity              & $Z_{w,\textnormal{ref}}^*$         & Sets metallicity at which transition from high- to low-energy galactic winds occurs. Sampled between 0.0005 and 0.008 with fiducial value 0.002.                                                                      \\ \hline
WindEnergyReductionExponent                  & $\gamma_{w,Z}^*$                   & Controls abruptness in metallicity of transition between high- and low-energy galactic winds. Sampled linearly between 1 and 3 with fiducial value 2.                                                                                 \\ \hline
WindDumpFactor                               & $1-\gamma_w^*$                     & Fraction of metals in SF cell getting ejected into a galactic wind that are deposited in neighboring SF cells prior to ejection. Sampled linearly between 0.2 and 1 with fiducial value 0.6.                                          \\ \hline
SeedBlackHoleMass                            & $M_\textnormal{BHseed}$            & Mass of seed SMBHs. Sampled between $2.5\times 10^5$ and $2.5\times 10^6 M_\odot$ with fiducial value $8\times10^5 M_\odot$.~                                                                                         \\ \hline
BlackHoleAccretionFactor                     & $A_\textnormal{Bondi}$           & Normalization factor for the Bondi rate for accretion onto SMBHs. Sampled between 0.25 and 4 with fiducial value 1.                                                                                                   \\ \hline
BlackHoleEddingtonFactor                     & $A_\textnormal{Edd}$             & Normalization factor for the limiting Eddington rate for accretion onto SMBHs. Sampled between 0.1 and 10 with fiducial value 1.                                                                                      \\ \hline
BlackHoleFeedbackFactor                      & $\epsilon_{f,\textnormal{high}}$ & Normalization factor for the energy in AGN feedback per unit accretion rate in the high-accretion state. Sampled between 0.025 and 0.4 with fiducial value 0.1.                                                       \\ \hline
BlackHoleRadiativeEfficiency                 & $\epsilon_r$                     & Fraction of the accretion rest-mass that is released in the accretion process. Sampled between 0.05 and 0.8 with fiducial value 0.2.                                                                                  \\ \hline
QuasarThreshold                              & $\chi_0$                         & Eddington ratio that serves as the threshold between the low- and high-accretion states of AGN feedback. Sampled between 0.000063 and 0.063 with fiducial value 0.002.                                                \\ \hline
QuasarThresholdPower                         & $\beta$                          & Power-law index of scaling of the low- to high-accretion state threshold with BH mass. Sampled linearly between 0 and 4 with fiducial value 2.   \\ \hline                                                                         
\end{tabularx}
\end{table*}

\begin{table*}\caption{Summary of the galactic feedback parameters in the CAMELS-Simba suite. Values are sampled logarithmically unless otherwise stated. More details about the parameters can be found in \citet{Ni:2023}.}\label{tab:SIMBAsummary}
\centering
\begin{tabularx}{\textwidth}{l|c|p{0.6\textwidth}}
\hline
\multicolumn{1}{c}{Name from Ni et al. 2023} & \multicolumn{1}{c}{Symbol} & \multicolumn{1}{c}{Description}  \\ \hline \hline
$A_{SN1}$                          & -                        & Normalization factor for the mass loading of galactic winds. Sampled between 0.25 and 4.0 with fiducial value 1.  \\ \hline
$A_{SN2}$                         & -                       & Normalization factor for the speed of galactic winds. Sampled between 0.5 and 2 with fiducial value 1.  \\ \hline
$A_{AGN1}$                          & -                       & Normalization factor for the momentum flux of kinetic AGN-driven overflows. Sampled between 0.25 and 4 with fiducial value 1.   \\ \hline
$A_{AGN2}$            & -                       & Normalization factor for the speed of AGN-driven outflows in jet mode. Sampled between 0.5 and 2 with fiducial value 1.                                                            \\ \hline
SfrCritDens                              & $n^*_\textnormal{crit}$                          & Gas number density threshold for SF. Sampled between 0.02 and 2 cm$^{-3}$ with fiducial value 0.2 cm$^{-3}$.                                                                                                     \\ \hline
SfrEfficiency                           & $\epsilon_*$                        & SF efficiency per free fall time. Sampled between 0.01 and 0.04 with fiducial value 0.02.                                                                                         \\ \hline
ISMJeansFac                                    & $A_\textnormal{ISMngb}$                         & Determines the level of artificial pressurization in the ISM. Sampled between 0.25 and 4 with fiducial value 1.                                                                             \\ \hline
WindTravTime                          & $t_{rc}^*$        & Maximum amount of time that galactic winds can propagate before recoupling in units of the Hubble time. Sampled between 0.2 and 0.002 with fiducial value 0.02                                                                                     \\ \hline
WindDensFac                         & $n_{rc}^*$                         &  Defines gas number density threshold below which decoupled galactic winds are forces to recouple. Sampled between 0.1 and 0.001 cm$^{-3}$ with fiducial value 0.01 cm$^{-3}$.                                                                                          \\ \hline
WindColdTemp                    & $T_{\textnormal{cold}}^*$             & Defines the temperature assumed for the cold phase of galactic wnds. Sampled between 100 and $10^4$ K with fiducial value 1000 K.                                                                      \\ \hline
WindHotFrac                        & $A_\textnormal{hot}^*$                    & Fraction of wind particles ejected in the hot phase. Sampled linearly between 0 and 0.6 with fiducial value 0.3.  \\ \hline
WindVelSlope                                  & $\gamma_\textnormal{v}^*$         & Power-law dependence of the galactic wind velocity on the circular velocity of the galaxy. Sampled linearly between -0.38 and 0.62 with fiducial value 0.12.                                                                                           \\ \hline
AGBWindHeatVel                    & $v_\textnormal{AGB}^*$                        & Velocity of winds from AGB stars which are assumed to thermalize with the ambient ISM. Sampled between 25 and 400 km s$^{-1}$ with fiducial value 100 km s$^{-1}$.                                                                     \\ \hline
BHSeedMass              & $M_{\textnormal{BHseed}}$         & Mass of the SMBH seeds sampled between $10^3$ and $10^5 h^{-1}M_\odot$ with fiducial value $10^4 h^{-1}M_\odot$.                                                                      \\ \hline
BHSeedRatio                 & $A_{\textnormal{BHseed}}$                   & Normalization factor for the minimum galaxy stellar mass required for BH seeding. Sampled between $3\times 10^4$ and $3\times10^6$ with fiducial value $3\times10^5$.                                                                               \\ \hline
BHAccrFac                               & $A_{\textnormal{BHaccr}}$                     &  Normalization factor for the BH growth rate. Sampled between 0.25 and 4 with fiducial value 1.                                        \\ \hline
BHAccrMaxR                            & $R_{\textnormal{BHkernel}}$            & Maximum size of the BH kernel used with searching for neighboring gas elements. Sampled between 2 and 8 $h^{-1}$kpc with fiducial value 4 $h^{-1}$kpc.                                                                                     \\ \hline
BHAccrTempThr                     & $T_\textnormal{AccrThr}$           & Temperature threshold above which gas accretes onto the BH at the Bondi rate. Sampled between $10^4$ and $10^6$ K with fiducial value $10^5$ K.                                                                                                   \\ \hline
BHEddingtonFac                    & $A_\textnormal{torque}$             & Normalization factor for limiting Eddington rate of cold gas accretion. Sampled between 0.75 and 12 with fiducial value of 3.                                                                                     \\ \hline
BHNgbFac                     & $A_\textnormal{BHngb}$ & Sets desired effective number of gas resolution elements within the BH kernel. Sampled linearly between 2 and 6 with fiducial value 4.                                                      \\ \hline
BHRadiativeEff                & $\eta$                     & Radiative efficiency of the BH accretion disk. Sampled between 0.025 and 0.4 with fiducial value 0.1.                                                                                  \\ \hline
BHJetTvirVel                              & $v_\textnormal{BHvir}$                         & Sets the outflow velocity above which AGN jets are heated to the virial temperature of the halo. Sampled between 500 and 8000 km s$^{-1}$ with fiducial value 2000 km s$^{-1}$.                                                \\ \hline
BHJetMassThr                         & $M_\textnormal{BHjet}$                          & BH mass threshold above which BHs are allowed to transition into the jet feedback mode. Sampled between $4.5\times10^6$ and $4.5\times10^8 M_\odot$ with fiducial value $4.5\times10^7 M_\odot$.   \\ \hline                                                                                    
\end{tabularx}
\end{table*}

\bibliography{bib}

\end{document}